\documentclass[aps, pra, twocolumn, showpacs,nofootinbib,superscriptaddress,longbibliography]{revtex4-1}

\usepackage{psfrag,graphicx}
\usepackage{dcolumn}
\usepackage{bm}
\usepackage{amsfonts,amssymb,amsmath,amsthm}
\usepackage{lipsum}
\usepackage{dsfont}
\usepackage{physics}
\usepackage{comment}
\usepackage[justification=raggedright,font=small,labelfont=bf,format=plain]{caption}
\usepackage{subcaption}
\usepackage[usenames, dvipsnames]{color}
\usepackage{xcolor}
\usepackage{comment}
\usepackage{mathtools}
\usepackage{todonotes}
\usepackage{tikz,siunitx}
\usetikzlibrary{shapes.geometric,shapes.symbols}

\newcommand{\Imperial}{Blackett Laboratory, Imperial College London, SW7 2AZ, United Kingdom}
\newcommand{\somap}{School of Mathematical and Physical Sciences, Macquarie University,  NSW 2109, Australia}
\newcommand{\Sydney}{ARC Centre of Excellence for Engineered Quantum Systems, Macquarie University,  NSW 2109, Australia}
\newcommand{\Erlangen}{Physics Department, Friedrich-Alexander Universit\"{a}t of Erlangen-Nuremberg, Staudtstr. 7, 91058 Erlangen, Germany
}
\newcommand{\Dresden}{Helmholtz-Zentrum Dresden-Rossendorf, Bautzner Landstraße 400, 01328 Dresden, Germany}

\begin{document}

\title{Creation and manipulation of surface code defects with quantum optimal control}
\date{2024}

\author{Omar Raii}\affiliation{\Imperial}\affiliation{\somap}
\author{Anirban Dey}\affiliation{\somap}\affiliation{\Sydney}
\author{Florian Mintert}\affiliation{\Imperial}\affiliation{\Dresden}
\author{Daniel Burgarth}\affiliation{\somap}\affiliation{\Erlangen}

\begin{abstract}
The surface code is a spin-1/2 lattice system that can exhibit non-trivial topological order when defects are punctured in the lattice and thus can be used as a stabiliser code. The protocols developed to create defects in the system have previously relied on adiabatic dynamics. In this work we use techniques of quantum optimal control to overcome the requirement for adiabaticity and achieve defect creation and implemention of other important operations required for quantum computation at much faster timescales. 

\end{abstract}

\maketitle

\section*{Introduction}

Of the many varied proposals for the hardware of systems for quantum computation, topological quantum systems, that is, systems with non-trivial topological order, present numerous advantages from a structural point of view \cite{Cesare2015AdiabaticComputing,Brennen2007WhyAnyons,Nayak2008,Lahtinen2017AComputation,Sarma2015}. The inherently non-local nature of topologically encoded logical qubits as well as the  nature of quantum logical gate implementation through braiding implies an inherent robustness to local noise \cite{Nakamura2020DirectStatistics,Manna2022AnyonHamiltonian}.

The surface code, a generalisation of the toric double quantum code \cite{Kitaev1997} that does not require periodic boundary conditions, is a spin-qubit lattice that provides an appealing example of a physical system that can exhibit non-trivial topological order. It can be defined by a Hamiltonian comprising primarily of four-body interactions known as stabilisers. When certain stabilisers are removed by a process known as `puncturing' \cite{Cesare2015AdiabaticComputing,Brown2017PokingCode}, defects are created in the code wherein quantum information may be encoded.

As with other examples of topological systems which have shown potential in terms of suitability for quantum computation, the realisation of anyons and anyon-like defects has relied on using adiabatic dynamics \cite{Lahtinen2011InteractingModel,Cesare2015AdiabaticComputing} and therefore requires long timescales. Such requirements amount to a limitation on the experimental realisation of the system given the necessity of implementing quantum algorithms quickly with respect to decoherence times. In this work we seek to overcome this limitation via the use of time-optimal quantum control.

The techniques of quantum optimal control have been shown to successfully effect desired dynamics while bypassing adiabatic, therefore slow, timescales \cite{He2016EfficientDriving,Zhou2017AcceleratedSystem,Xu2019BreakingGeodesics,Wang2012AdiabaticControl,Simsek2021QuantumInvariant,Raii2022ScalableModel}. Quantum control in general relies on determining specific chosen time-dependent Hamiltonians in order to implement desired dynamics \cite{DAlessandro2008,Chen2021QuantumIntroduction}. We seek to use numerical quantum optimal control to successfully carry out operations required for encoding and processing logical qubits in the surface code within timescales that are much shorter than those which would be required by adiabatic protocols.

One potentially significant obstacle to numerical solving of dynamics in lattice systems is the large Hilbert space size when operating with many spins. Our work presents a method where conserved quantities are used to break down the control problem into many smaller control problems which are each solvable separably.

The problem then essentially becomes an ensemble control problem, where we want each system to achieve a different target unitary using a single control pulse simultaneously. Each individual system in the ensemble is made up of a drift Hamiltonian and a control Hamiltonian specific to the drift Hamiltonian. We employ a gradient based optimization algorithm, to find a control pulse that minimizes the average fidelity error across the ensemble. We aim for the average error to get below a predefined target error, $\varepsilon$. For each system in the ensemble, we calculate the individual fidelity against its corresponding target unitary using a phase-sensitive fidelity function.

Section I of this work presents a brief overview of the surface code, followed by a description in Section II of how adiabatically changing Hamiltonians allows for defects in the surface code to be created and manipulated, as described in more detail in the literature \cite{Cesare2015AdiabaticComputing}. In Section III we describe in detail the quantum control methods used and present our optimisation results for the four major operations required for encoding and manipulating quantum information in the surface code. For three of the four operations this is comprised of numerical optimisation while for the fourth operation, known as surface code detachment, the aforementioned technique of using conserved quantities is applied and this is described in detail in Section IV.

For those familiar with the surface code and adiabatic procedures on it, Sections I and II are for review purposes and are not essential.

\section{surface code}

As with the toric code \cite{Kitaev1997}, the surface code lattice comprises of a large number of spin-1/2 degrees of freedom on the edges of a square lattice as shown in Fig.~\ref{fig:ZZdefects}. Stabiliser operators are defined by four-body Pauli $X$ interactions on vertices and Pauli $Z$ operators on plaquettes. The vertex and plaquette operators, $A_v$ and $B_p$ are respectively defined on a vertex $v$ and plaquette $p$ as
\begin{align}
    A_v : &= X_{v_1,v_2,v_3,v_4} \\
     B_p : &= Z_{p_1,p_2,p_3,p_4}
\end{align}
where $X_{v_1,v_2,v_3,v_4} := X_{v_1} X_{v_2} X_{v_3} X_{v_4}$ and $Z_{p_1,p_2,p_3,p_4} := Z_{p_1} Z_{p_2} Z_{p_3} Z_{p_4}$.

Lattices may include `rough sides' where plaquette operators on the edge are defined only with three-body terms, as well as `smooth sides' where vertex operators on such edges are defined with three-body terms. If both smooth \textit{and} rough sides are present then for two of the four corners of the lattice (in the example shown in Fig.~ \ref{fig:ZZdefects} these would be the top right corner and the bottom left corner) there is defined for each a corresponding \textit{two-body} stabiliser, either a vertex or plaquette operator.

With the stabilisers defined, one can see that as a plaquette stabiliser $B_p$ and a vertex stabiliser $A_v$ can overlap only on an even number of qubits, all stabilisers must therefore commute with one another. Additionally each stabiliser squares to the identity. Defining the system Hamiltonian 
\begin{align}
    H_0 &= -\frac{\Delta}{2} \left( \sum_v A_v + \sum_p B_p \right)
    \label{eq:SurfaceCodeHam}
\end{align}
with $\Delta$ being the spectral gap, the ground state up to normalisation is then same as the toric code
\begin{align}
    \ket{g_0} &= \prod_v (\textbf{1}+ A_v) \ket{0}^{\otimes N}.
\end{align}
Unlike the toric code however, the ground state space is non-degenerate which would initially seems to make it unfeasible for use as a system for topological quantum computation. However the required degeneracies are introduced through `puncturing' of the lattice, an operation amounting to the removal of certain stabilisers from the Hamiltonian. Each puncture doubles the size of the ground state degeneracy.

\section{Hamiltonians, stabiliser formalism and adiabaticity}

Quantum information is encoded in the lattice via the stabiliser formalism whereby a logical qubit is defined by a set of stabiliser operators such that the qubit corresponds to a physical state which is simultaneously in a $+1$ eigenstate of the entire stabiliser set. 
If we were to start with a full surface code lattice and implement a defect, that is we remove either a plaquette ($Z$ defect) or a vertex ($X$ defect) stabiliser, we would create a code whereby the logical gate $\bar{Z}$ corresponds to a string of Pauli $Z$ operations on the lattice around the defect. The logical gate $\bar{X}$ would then correspond to a string of Pauli $X$ operations from the smooth defect to a smooth edge. The aim therefore is, when beginning in the ground state of an initial Hamiltonian $H_i$, to end up in the corresponding state of a final Hamiltonian $H_f$ which amounts logically to an encoding of a specific logical quantum state. Each Hamiltonian comprises of the stabiliser set of a different configuration of logical quantum information. The deformation of these codes is studied in the literature as physically being implemented through `adiabatic dragging' \cite{Cesare2015AdiabaticComputing,Bacon2009AdiabaticTeleportation,Bacon2010AdiabaticComputing} between the initial and final Hamiltonians of each particular procedure. (see \cite{Cesare2015AdiabaticComputing} for an in-depth description).

There are four major operations that can be implemented in order to create defects in the surface code, encode logical qubits within them, and manipulate their location or grow them in size to give the encoded qubits further protection. 

\subsection{Defect creation}

\begin{figure}[h]
\centering

  \centering
  \includegraphics[width=\linewidth]{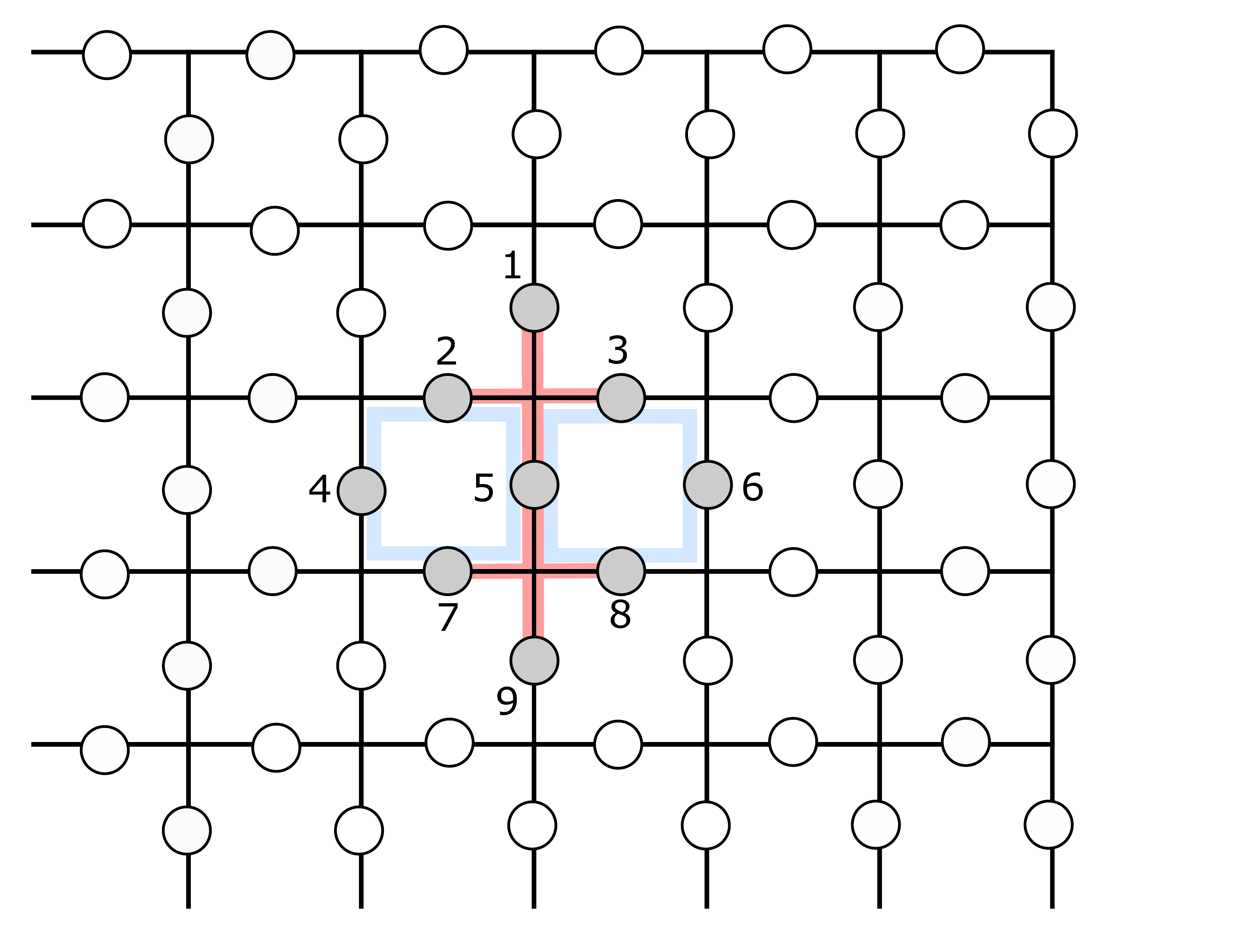}
  \captionof{figure}{Depiction of the spin-qubits acted on during defect creation. Only 9 spin-qubits are affected. The two plaquettes $B_{p_1} = Z_{2,4,5,7}$ and $B_{p_2} = Z_{3,5,6,8}$ are removed from the Hamiltonian and turn from stabilisers into logical gates ($\bar{Z}$) on the encoded qubits. The upper vertex $v_1$ consists of qubits 1,2,3 and 5 while the lower vertex $v_2$ corresponds to qubits 5,7,8 and 9.}
  \label{fig:ZZdefects}
\end{figure}

Described here is the creation of a pair of smooth ($Z$-type) defects and logical $\ket{\bar{0}}$ states in the surface code. The creation of rough defects is an analogous procedure, with the roles of $X$ and $Z$ interactions reversed. Labelling of spins throughout the work will follow a convention of numbering sites from rows going top to bottom and from left to right within each row, such as that which is shown in Fig.~\ref{fig:ZZdefects}.

The procedure consists of starting with the Hamiltonian described in Equation (\ref{eq:SurfaceCodeHam}) as an initial Hamiltonian and transforming it into a final Hamiltonian where particular stabilisers are removed.  
Given that all of the vertex and plaquette stabilisers which are not involved in the defect creation commute with all the terms in the initial and final Hamiltonian, the Hilbert space of the dynamics of interest reduces to $2^9$ and thus it suffices to consider the initial and final Hamiltonians
\begin{align}
    \tilde{H}_i &=  - \frac{\Delta}{2} ( X_{1,2,3,5} + X_{5,7,8,9} +   Z_{2,4,5,7} + Z_{3,5,6,8} ) \\
    \tilde{H}_f &= - \frac{\Delta}{2} (X_{1,2,3} + X_{7,8,9} ) - \frac{\Delta}{2} X_5
\end{align}
where the spectral gap $\Delta$ remains unaltered despite the system size reduction.

Qualitatively, this amounts to turning off the plaquette stabilisers $Z_{2,4,5,7}$ and $Z_{3,5,6,8}$ from the Hamiltonian, which would then define these operators as \textit{logical gates} $\bar{Z}$  of the two logical qubits encoded in the ground state of the final Hamiltonian.  When the two plaquette stabilisers are removed in addition the overlapping vertex stabilisers $X_{1,2,3,5}$ and $X_{5,7,8,9}$ are modified from four-body operators to three-body ones. For a smooth defect, a logical $X$ gate is then simply a string of Pauli $X$ operators defined on a string traversing the geometrically dual lattice starting from the defect and ending at a smooth edge. These two $\bar{X}$ gate strings can be made to overlap on just a single spin such as spin 5 as labelled in Fig.~\ref{fig:ZZdefects} and this accounts for the single $X_5$ term in the final Hamiltonian

%{Two adjacent plaquette operators can be promoted from stabilisers into logical $Z$ gates by removing them from the Hamiltonian. Two such plaquettes are highlighted in blue in Fig.~\ref{fig:ZZdefects}. Along with the requirement for the two logical qubits which are to be created to have corresponding $\bar{Z}$ gates, a further requirement is for each qubit to have corresponding $\bar{X}$ gates. For a smooth defect, a logical $X$ gate is simply a string of Pauli $X$ operators defined on a string traversing the dual lattice starting from the defect and ending at a smooth edge (this dual lattice is simply a geometric spatial description corresponding to a lattice whose edges bisect the edges of the original lattice through the positions of the original lattice's spin qubits. The dual lattice is not to be confused with the reciprocal lattice which rather exists in momentum space). These two $\bar{X}$ gate strings can be made to overlap on just a single qubit such as spin-qubit 5 as labelled in Fig.~\ref{fig:ZZdefects}. When the two plaquette stabilisers are removed in addition the overlapping vertex stabilisers are modified from four-body operators to three-body ones.}

In summary, in defect creation the system begins in the unique ground state of $H_0$ and ends up in the logical state $\ket{\bar{0}}$ for each of the two defects that have been created, each of which have corresponding $\bar{Z}$ and $\bar{X}$ gates.

\subsection{Deformation of defects}

While logical qubits can be encoded onto defects, these defects need to be moved around on the lattice in order to implement logical gates. A typical example is the CNOT gate which is implemented by the braiding of one defect around another. It is also desirable for the defect to be grown so that the defect spans multiple plaquettes (or vertices) which gives greater protection against unwanted logical gate errors. The deformation of defects can involve four possible scenarios depending on the number of interior spins, from 1 to 4, within a defect that need to be acted on.

Each scenario differs slightly from the others and we describe the first scenario here as a concrete example. This operation only involves interaction with 7 qubits with the aim of growing a smooth defect such that it transforms from being localised on a two-by-two square of plaquettes to a defect covering five plaquettes. After labelling with the qubits following the same convention as previously mentioned, the initial and final Hamiltonians are given respectively as
\begin{align}
    H_i &= -\frac{\Delta}{2} (Z_{3,5,6,7} + X_{2,3,5} + X_{1,3,4,6})
\end{align}
and
\begin{align}
    H_f &= -\frac{\Delta}{2} ( X_3 + X_{2,5} + X_{1,4,6}).
\end{align}

The other three scenarios are conceptually similar except that they only involve interactions with 8, 8 and 4 spins respectively. 

\begin{figure}[h]
\centering

  \centering
  \includegraphics[width=\linewidth]{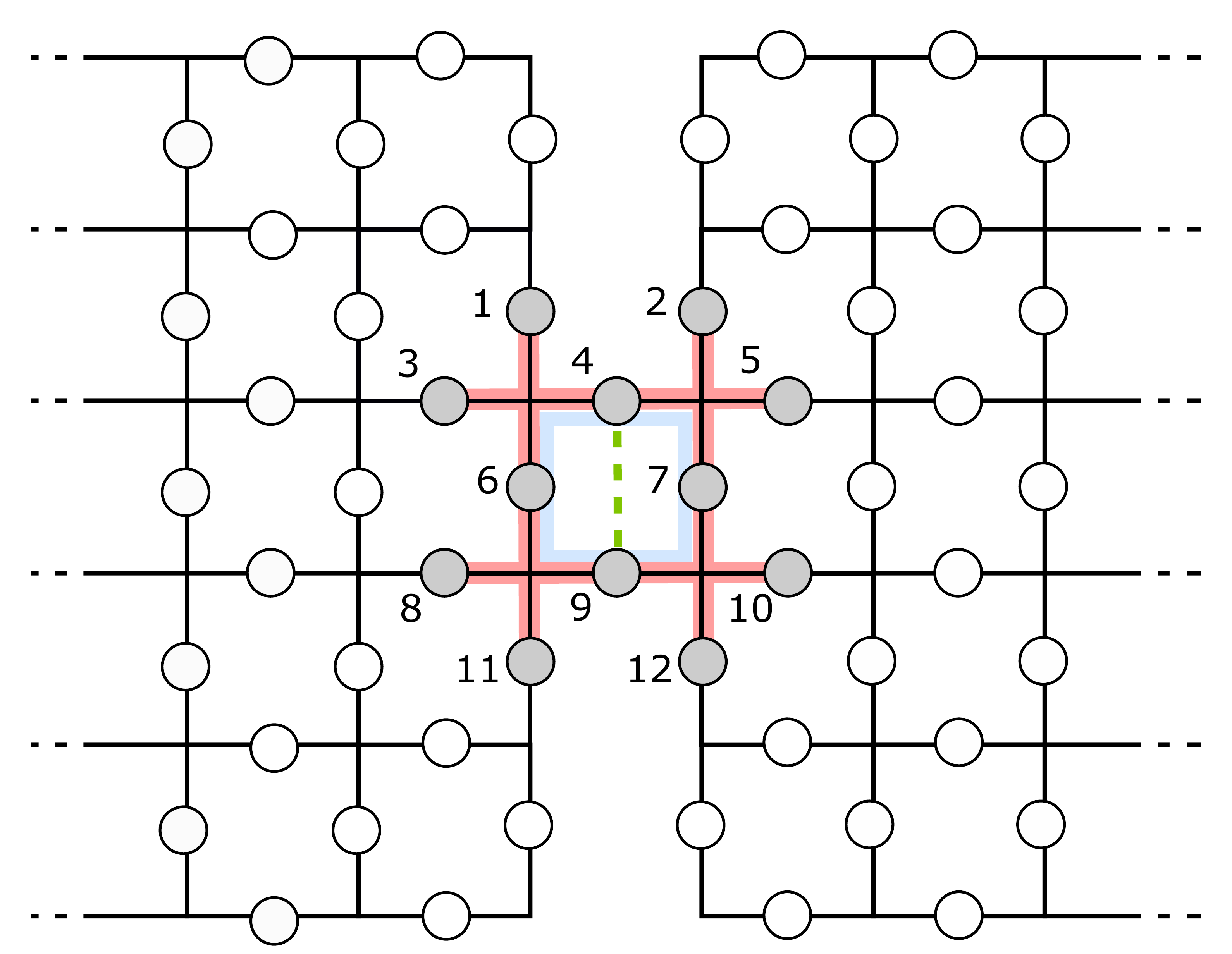}
  \captionof{figure}{The areas are to be separated through the removal of stabiliser connecting them. The four, surrounding, four-body vertex operators then become three-body terms.}
  \label{fig:detachmoat}
\end{figure}

Multi-step operations such as state injection (described below) require detachment and attachment of surface code areas containing defects. Looking at the example of detachment, the removal of an area is shown in Fig.~\ref{fig:detachmoat}; this amounts to the removal of a plaquette stabiliser operator on the plaquette connecting the two regions.

Explicitly this corresponds to starting with an initial Hamiltonian
\begin{align}
\begin{split}
    H_i = -\frac{\Delta}{2} (&X_{1,3,4,6} + X_{2,4,5,7} + X_{6,8,9,11} \\
    + &X_{7,9,10,12} + X_{4,9} + Z_{4,6,7,9} ) 
\end{split}
\label{eq:hiDetachment1}
\end{align}
and ending with a final Hamiltonian
\begin{align}
    \begin{split}
    H_f = -\frac{\Delta}{2} (&X_{1,3,6}+ X_{2,5,7} + X_{6,8,11} \\ + &X_{7,10,12} + X_4 + X_9 )
    \end{split}
    \label{eq:hfDetachment2}
\end{align}
Here the initial Hamiltonian $H_i$ is the negative of the sum of the four body terms, one of which is the plaquette stabiliser that is to be turned off, as well as a two-body check term $X_4 X_9$. The final Hamiltonian $H_f$ is then comprised of each four-body stabiliser turned into a three-body stabiliser, the plaquette stabiliser removed, and the two-body check term replaced by two single body terms. 

This operation requires all overlapping vertex stabiliser operators, of which there are four, to be modified from four-body terms into three-body terms, such that if we wish to remove the plaquette operator and modify the vertex operators in one go, this would affect 12 spin and so the Hilbert space size where the dynamics are implemented is large compared with the other operations. 

\subsection{State injection}
This scenario involves turning the logical states that have been encoded upon the defects, whether $\ket{\bar{0}}$ or $\ket{\bar{+}}$, into various fiducial states such as the canonical example of $\bar{T}\ket{\bar{+}}$ where $T$ is the phase gate such that $T^4 = Z$.

This comprises several steps, including: i) Creating a pair of rough defects in an eigenstate of $X$; this procedure is analagous to that of creating smooth defects described previously. ii) Ensuring that the defect is in a region which is smoothly detached from the rest of the surface code. This involves the detachment procedure described in the detachment/attachment section. iii) Adiabatically or otherwise `injecting' the desired logical state onto the spin qubit located between the pair of defects. This simply means evolving the state of the spin qubit from $\ket{\psi}$ to $TX\ket{\psi}$.  iv) Deforming the pair of defects  from rough to smooth by turning off the two adjoining (three-body) plaquette stabilisers and turning on the vertex stabilisers.

Out of these, the first procedure which is distinct from the other three operations is the actual state injection which is carried out on a single qubit in the interior of a pair of rough defects. The second such procedure consists of turning on the adjoining vertex operators and turning off the overlapping three-body plaquette operators. Hence we can define initial and final Hamiltonians 
\begin{align}
    H_i &= - \frac{\Delta}{2} \left( Z_{1,2,3} + Z_{7,8,9}  \right) \\
    H_f &= -\frac{\Delta}{2} \left( X_{2,4,5,7} + X_{3,5,6,8} \right)
\end{align}
respectively.
Solving the dynamics of this second procedure is more numerically challenging than for the first procedure as the latter takes place within a $2^9$-dimensional Hilbert space. 

\section{Quantum optimal control}

\subsection{Defining the control problem}

Considering the implementation of these operations as quantum control problems involves reformulating the initial and final Hamiltonians into a more general time-dependent Hamiltonian with a control $f(t)$. Typically such a time-dependent Hamiltonian includes a time-independent part known as the drift Hamiltonian and a control function $f(t)$ which is coupled to a control Hamiltonian. Analogously, two Hamiltonians are chosen 
\begin{align}
    H_1 &= \tilde{H}_i \\ 
    H_2 &= \tilde{H}_f - \tilde{H}_i
\end{align}
so that we have the control problem defined as
\begin{align}
    H(t) = H_1 + f(t) H_2.
\end{align}
In the adiabatic regime $f(t)$ may simply be a linear ramp with a very small slope such that $f(0)=0$ and $f(T)=1$ for $T$ being long timescale, thereby ensuring adiabatic dynamics. We call such a pulse a linear ramp $f_a(t)$

We define an initial state $\ket{g_1}$ being a ground state of $H_1$ and this in turn defines a target state $\ket{g_\text{targ}}$ by
\begin{align}
    \ket{g_\text{targ}} = U(f_a(T)) \ket{g_1}
\end{align}
where adiabaticity ensures that $\ket{g_\text{targ}}$ is a ground state of the final Hamiltonian $H_1 + H_2$. This allows for the definition of a target fidelity, as a function of the control pulse 
\begin{align}
    \mathcal{F} (U(f,t) )= \left| \bra{g_\text{targ}} U(f(t)) \ket{g_1} \right|^2.
\end{align}
The closely associated target infidelity $\mathcal{I}= 1-\mathcal{F}$ is the figure of merit which is often sought to be minimised.

For some other operations, where the goal is to implement desired dynamics in full rather than specific state transfer, a similar unitary \textit{gate fidelity} defined by
\begin{align}
    \mathcal{F}(U) = \frac{1}{d} \left| \Tr(U_\text{targ}^\dagger U ) \right|
    \label{eq:matrixfidelity}
\end{align}
may instead be considered.

In order to optimise for infidelities, we use a gradient ascent pulse engineering algorithm (GRAPE) \cite{Khaneja2005OptimalAlgorithms,Rowland2012ImplementingPracticalities}. This procedure uses piecewise constant control pulses where, after an initial random control is picked,  the parameters such as pulse magnitude are varied in order for infidelity to go down in the control space until acceptable minima are found. The direction of descent in control space is determined by second derivatives of the infidelity which are calculated using the limited memory version of the Broyden–Fletcher–Goldfarb–Shanno method (BFGS) \cite{Broyden1970TheConsiderations,Fletcher1970AAlgorithms,Goldfarb1970AMeans,Shanno1970ConditioningMinimization,Liu1989OnOptimization}. This method uses an estimate of the Hessian matrix of second derivatives of the figure of merit and allows for a parabola to be used for approximating the control space and descending towards local minima.

GRAPE is used to find appropriate piecewise constant control pulses and choose the best fidelity for a given time period. Numerous parameters may be varied such as initial pulse guess and number of pulse time steps.

\subsection{Non-adiabatic optimisation}

\begin{figure}[h]
  \centering
  \includegraphics[width=\linewidth]{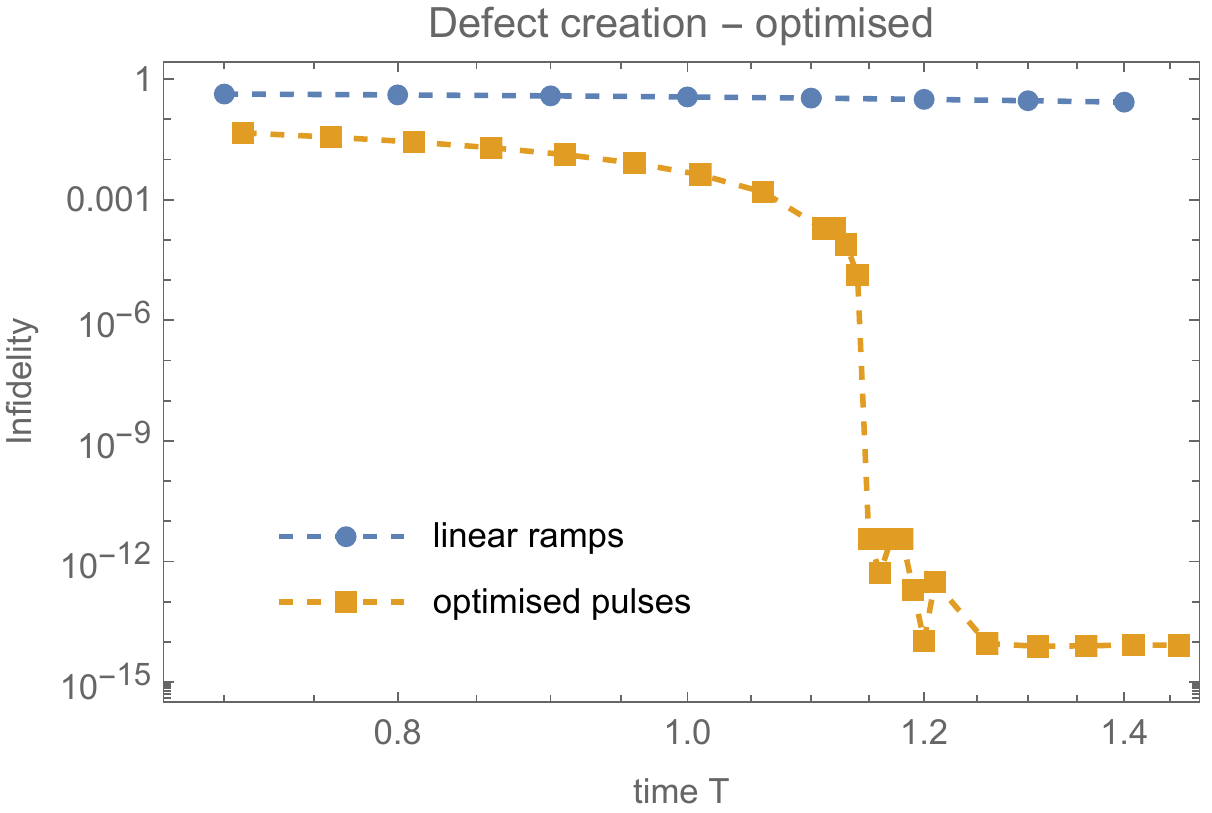}
  \captionof{figure}{Defect creation with optimised non-linear control pulses. The timescale for defect creation with high fidelity is around $T= 1.15$. Time units are $\Delta^{-1}$. The optimised fidelities are shown in orange as compared with the infidelity for a linear ramp at the corresponding time, shown in blue. This is an orders of magnitude improvement compared with using linear adiabatic pulses.}
  \label{fig:optimisedCreation}
\end{figure}

The result of our optimisations on three of the four major surface code operations, presented in Fig.~\ref{fig:optimisedCreation} and Table~\ref{table:droptimetable}, indicate that quantum optimal control can indeed allow for an improvement of multiple orders of magnitude in timescale required for achieving the appropriate target states of the operations. 

\begin{figure}[h]
\centering

  \centering
  \includegraphics[width=\linewidth]{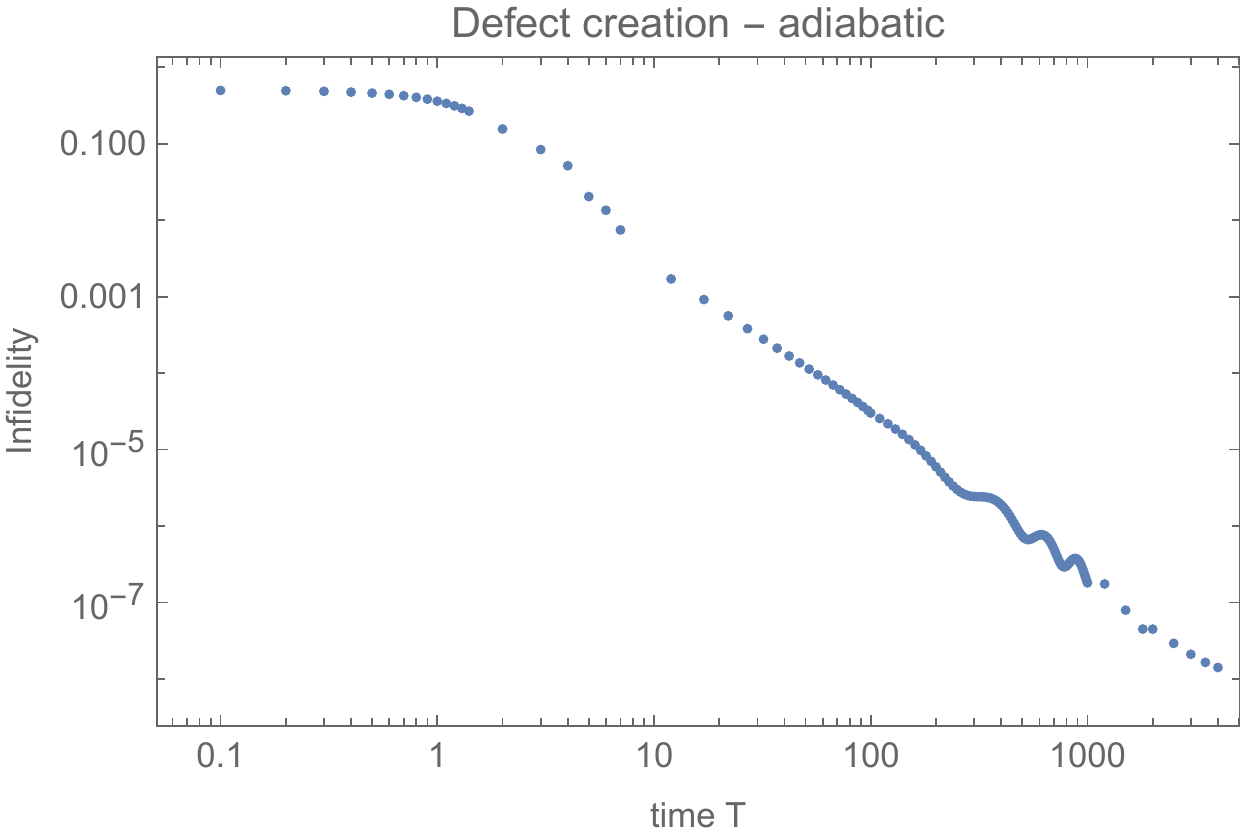}
  \captionof{figure}{Defect creation with linear control pulses. Fidelity when using linear ramps only becomes good at long timescales, where adiabaticity is achieved. Time units are $\Delta^{-1}$.}
  \label{fig:adiabaticCreation}
\end{figure}

In Fig.~\ref{fig:adiabaticCreation} are presented the state infidelities between the ideal target state, which corresponds to defect pair creation on the surface code, and the state achieved by using a linear control that goes from $0$ to $1$ in time $T$. Here as with elsewhere in this work timescales are understood to be in units of $\Delta^{-1}$ where $\Delta$ is the spectral gap in the initial Hamiltonian $H_1$. The logarithmic plot displays the improving fidelity that is achieved with linear controls of longer timescales. In particular, when $T<1$ there is a comparatively high infidelity that stays effectively constant whereas for timescales longer than this, the infidelity improves steadily. Considering an infidelity of $10^{-7}$ as a standard for sufficiently good infidelity we see that this is achieved for timescales of $T \gtrsim
1000$. This is indeed consistent
with the spectral gap of $1$ in the units of $\Delta^{-1}$ as an effectively perfect fidelity is expected when adiabaticity is achieved which can only occur according to the adiabatic theorem when the timescale is much longer than the inverse of the spectral gap \cite{Avron1987AdiabaticEffect}.

In Fig.~\ref{fig:optimisedCreation} are presented the results of using optimisation of state infidelity to find non-linear controls that can achieve high fidelities. Good fidelities were achieved with only two time steps in the optimised pulses and this is compared with the  Each data point corresponds to the minimum of 1000 optimisation attempts with random initial pulse guess. The results show that successfully creating the defect pair state, that is, to achieve a good infidelity of $10^{-7}$ as mentioned previously, can be achieved at a timescale of $T \approx 1.15$. The earliest timescale at which effectively perfect fidelity is achieved hereafter is referred to as \textit{drop time}. %{Here we understand timescales to be in units of $\Delta^{-1}$ where $\Delta$ is the spectral gap in the initial Hamiltonian $H_1$.}
Furthermore we see the comparison between use of non-optimised linear control ramps and the optimised pulses when comparing fidelities (plotted in blue and orange respectively) as there is a marked improvement, particularly after the drop time where infidelities are shown to improve by over 10 orders of magnitude.

The results of carrying out the same optimisation procedure for defect creation and for two other operations, namely defect deformation and state injection, are qualitatively similar and are presented in Table. \ref{table:droptimetable}. The results show that there is a variance of the drop time but all operations achieve effectively perfect fidelity at times much smaller than the adiabatic timescale. An appropriate measure of this is comparison between optimised infidelity at drop time with the infidelity using a linear control pulse with same duration. Quantum control optimisation methods can be seen therefore to give an infidelity improvement generally on the order of 10 orders of magnitude.

\begin{table}[h]
\begin{center}
\begin{tabular}{ c|c|c|c|c|c } 
\hline
Operation & spins & $T_d$ & $N_\tau$ & $\mathcal{I}$ linear & $\mathcal{I}$ optimised  \\
\hline
Creation& 9& 1.15& 2& 0.3251& 5.13e-13 \\
Deformation 1& 7& 2.0& 4& 0.323& 5.50e-12\\
Deformation 2& 8& 3.2& 4& 0.379& 8.11e-13\\
Deformation 3& 8& 10.0& 12& 0.166& 2.41e-11\\
Deformation 4& 4& 5.0& 10& 0.927& 3.32e-12\\
State injection pt.1& 1& 2.4& 4& 0.177& 4.37e-14\\
State injection pt.2& 9& 1.3& 15& 0.358& 2.23e-12\\
\hline
\end{tabular}
\end{center}
\caption{Results of quantum control optimisation on three surface code operations, namely defect creation, defect deformation and state injection. Shown in the table are: number of spins acted on in the operation, drop time, minimum required number of time steps in optimised pulses, infidelity with linear pulse at time $T_d$, optimised infidelity.}
\label{table:droptimetable}
\end{table}

\section{Surface code detachment as a control problem}

The operation of detachment and attachment of surface code regions is sufficiently distinct from the other three operation that it requires an alternative approach if we wish to use quantum control to implement it without resorting to adiabaticity. Two spins within the lattice require removal to implement code detachment. It is therefore necessary for all overlapping vertex stabiliser operators (of which there are four; see Fig.~\ref{fig:detachmoat}) to be modified from four-body terms into three-body terms, and so in order to remove the plaquette operator and modify the vertex operators, 12 spins must be affected. The effective Hilbert space size of this operation therefore is much larger than for any of the other operations. Moreover, as the primary goal here is to find controls that mimic the behaviour of adiabatic dynamics no matter the initial state, unitary gate fidelity must be optimised rather than the simpler case of state fidelity.

Attempting to solve the dynamics in such a Hilbert space is more challenging than in previous examples and, as optimisation must be carried out many times, this therefore requires large computational resources. In order to overcome this obstacle, we exploit the fact that the Hamiltonian $H(t)=H_i + f(t) H_f$, where $H_i$ and $H_f$ are defined as in \ref{eq:hiDetachment1} and \ref{eq:hfDetachment2}, commutes with the operator $X_{1,2,3,5,8,10,11,12}$ for all time $t$. This means that $H(t)$ can be block-diagonalised into $2^8$ blocks each of which are of size $2^4$-by-$2^4$. The number of blocks correspond to the number of possible tuples of eigenvalues of the eight operators $X_j, $ which commute with $H(t)$ for all time, where $j\in \{ 1,2,3,5,8,10,11,12\}$.

Many of these blocks are in fact identical with each other due to the fact that the eigenvalues $x_j$ of the operators $X_j$ always come in the following multiplied pairs: $x_1 x_3, x_2 x_5, x_8 x_{11}$ and $x_{10} x_{12}$. Therefore only $2^4$ sub-systems will be distinct control problems. 

In practice however, arising from the fact that for some systems, the spectrums of the drift, control and corresponding target unitary, are the same, only 8 systems are distinct up to unitary equivalence. As a consequence, optimisation will only require consideration of 8 sub-systems, drastically reducing the difficulty of numerical optimisation.

This method allows us to achieve a similar result as with the other major operations on the surface code, where high fidelities are realised without resorting to long timescales or scaling of the Hamiltonian with large amplitude control pulses. 

\subsection{Reduced control problems and averaging phase-sensitive fidelity}

The full time-dependent Hamiltonian of the control problem can be reformulated in terms of a time-independent drift Hamiltonian $H_d$ and a time-dependent control $H_c$ such that $H(t) = H_d + f(t) H_c$. As $H(t)$ can be block diagonalised such that 
\begin{align}
    H(t) = \bigoplus_{j=1}^{2^8} \tilde{H}_j,
\end{align}
then the full dynamics of the system can be given as the direct sum of the dynamics induced by each separate smaller Hamiltonian, such that
\begin{align}
    U = \bigoplus_{j=1}^{2^8} \tilde{U}_j,
\end{align}
where $U$ is the dynamics that solves the Schr\"{o}dinger equation with Hamiltonian $H(t)$ and $\tilde{U}_j$ correspond to the dynamics of the reduced systems with Hamiltonians $\tilde{H}_j$.

In order to find the fidelity between a target dynamics and dynamics induced by Hamiltonian with a given time-dependent control function, we may consider the magnitude of the average of a phase-sensitive fidelity between the unitary dynamics of corresponding blocks. If we define a phase-sensitive fidelity as  
\begin{align}
    \tilde{\mathcal{F}} (U,V) &= \frac{1}{d} \Tr(U^\dagger V),
\end{align}
then the full fidelity can be given by
\begin{align}
    \mathcal{F} (U,V) &= \frac{1}{k} \left|\sum_{j=1}^k  \tilde{\mathcal{F}} (\tilde{U}_j , \tilde{V}_j) \right|.
\end{align}
As the dynamics are ultimately a function of a time-dependent control function, such that $U = U[f(t)]$, the aim is to use the GRAPE algorithm to find a control function that maximises the fidelity (or minimises the correspondingly defined infidelity) compared with a target unitary defined by adiabatic dynamics. This method effectively means attempting to optimise many small control problems with the same time-dependent control $f(t)$ at the same time.

\subsection{Numerical results}

\begin{figure}[h]
\centering

  \centering
  \includegraphics[width=\linewidth]{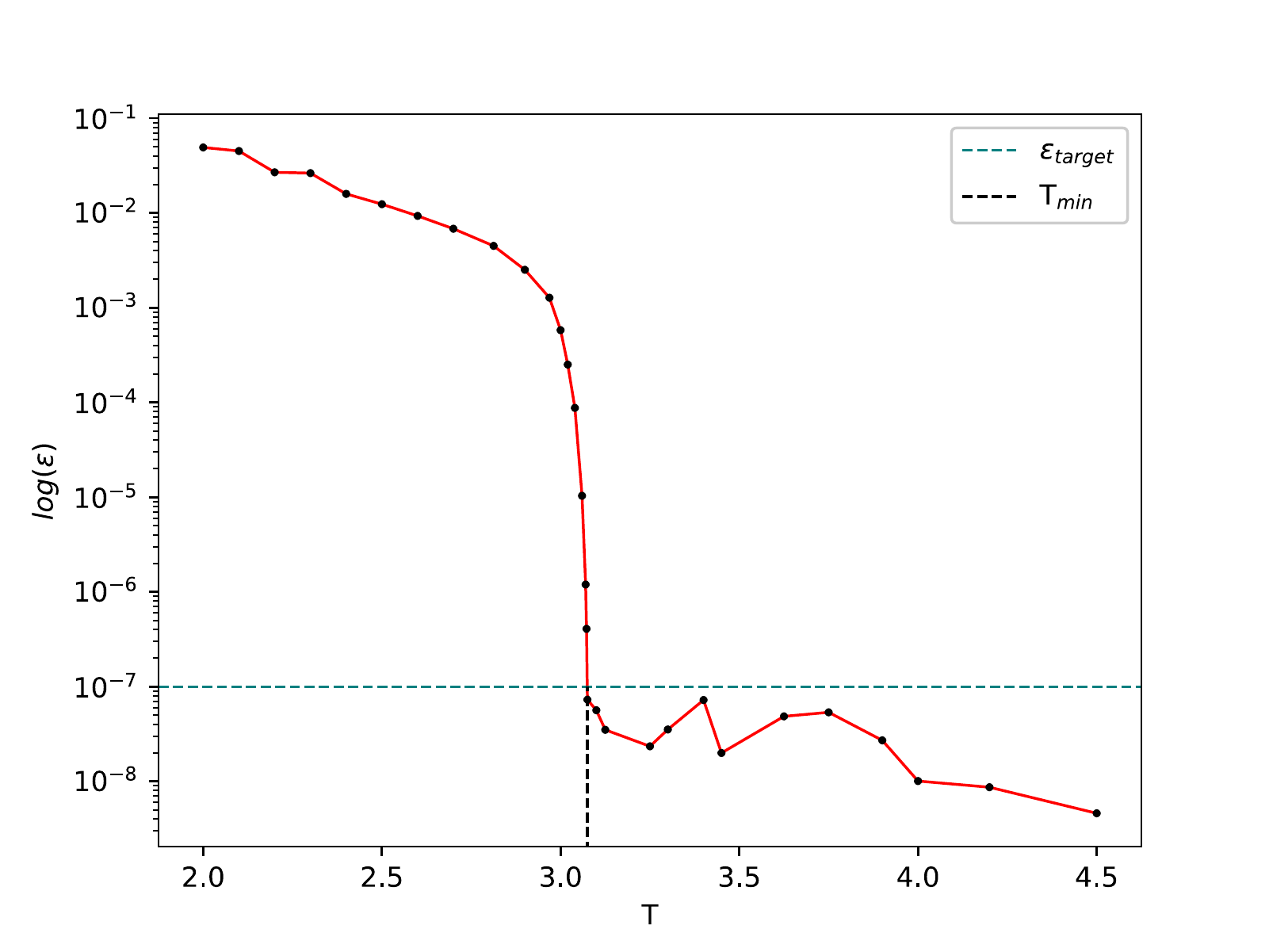}
  \captionof{figure}{Infidelities for the surface code detachment operation after optimisation for different timescales $T$. The teal line indicates the target infidelity of $10^{-7}$ and the black dashed line indicates the minimum time in which this target infidelity is achieved is $T_d = 3.075$ which is much lower than the adiabatic timescale.}
  \label{fig:minCtime}
\end{figure}

Fig. \ref{fig:minCtime} displays the total infidelity $\log(\varepsilon)$ for increasing evolution time $T$. The target infidelity of $10^{-7}$ was achieved at a time $T_d=3.075$, indicating an improvement of two orders of magnitude in the timescale compared to the adiabatic dynamics, which requires $T=1000$ to obtain an infidelity of $10^{-7}$ for the successful implementation of the surface code detachment operation.

\begin{figure}[h]
\centering

  \centering
  \includegraphics[width=\linewidth]{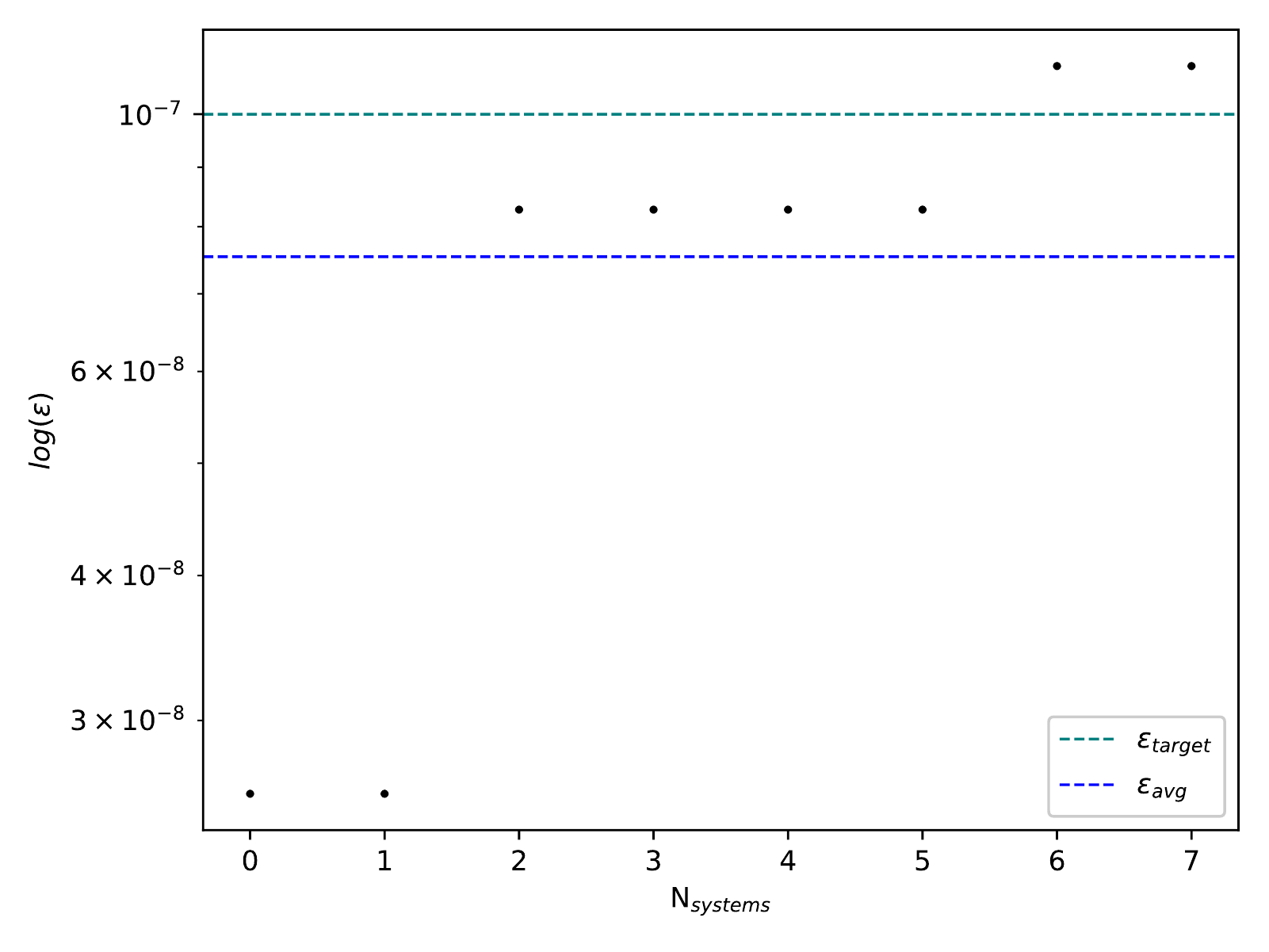}
  \captionof{figure}{Figure showing the 8 individual infidelities for the case of the optimised pulse for $T=3.075$. The teal dashed line indicates the average individual error in the target unitary matrix and the blue line indicates the average error we got after optimisation. This optimised average error lies below target error.}
  \label{fig:individErrors}
\end{figure}

For the case of the pulse leading to the target infidelity of $10^{-7}$ at the minimum time $T=3.075$, we compare the 8 individual infidelities of unitarily inequivalent sub-blocks within the full dynamics $\tilde{\mathcal{F}} (\tilde{U}_j, \tilde{V}_j)$, as shown in Fig. \ref{fig:individErrors}. The comparison indicates that infidelities never differ by more than one order of magnitude and that the average of the magnitudes of the individual infidelities is on the order of $10^{-7}$ after optimisation.
\section{Conclusion}

The results presented show that using quantum optimal control methods, it is possible to achieve the same high fidelities corresponding to successful encoding and processing of quantum information, as compared with using adiabatic dragging protocols. The methods presented allow for this to be achieved without resorting to the long timescales required by adiabaticity. 

Additionally it has been shown that it is possible to overcome the difficulty of implementing quantum control for a system where control is required on many spins within a lattice system, such as with the example of surface code detachment.  Successfully using commuting operators and conserved quantities, it has been shown how to break down the control problem into many smaller control problems and implement quantum optimal control on the small control problems at once. Optimising for an average phase-sensitive fidelitiy with the smaller problems, has led to achieving high total infidelity on the order of $10^{-7}$ at a much faster time scale compared to the ideal adiabatic dynamics.

\section*{Acknowledgements}
OR acknowledges funding from EPSRC Quantum Systems Engineering Hub and support from MQCQE. The authors would also like to thank Benjamin Brown and Lauritz van Luijk for their advice and useful discussion regarding the topics in the work. AD acknowledges funding from Sydney quantum academy (SQA) through primary PhD scholarship. DB acknowledges funding by the Australian Research Council (project numbers FT190100106, DP210101367, CE170100009).

\bibliographystyle{unsrt}
\bibliography{references%
  }

\end{document}